\documentclass[a4paper]{jpconf}
\usepackage{graphicx}
\begin{document}
\title{Low-temperature thermopower study of YbRh$_2$Si$_2$}

\author{S Hartmann, N Oeschler, C Krellner, C Geibel and F Steglich}
%\author{Niels Oeschler}
%\author{Cornelius Krellner}
%\author{Christoph Geibel}
%\author{Frank Steglich}

\address{Max Planck Institute for Chemical Physics of Solids, N\"othnitzer Str. 40, 01187 Dresden, Germany}

\ead{hartmann@cpfs.mpg.de}

\begin{abstract}
The heavy-fermion compound YbRh$_2$Si$_2$ exhibits an antiferromagnetic (AFM) phase transition at an extremely low temperature of $T_{\rm N}$ = 70~mK. Upon applying a tiny magnetic field of $B_c$ = 60~mT the AFM ordering is suppressed and the system is driven toward a field-induced quantum critical point (QCP). Here, we present low-temperature thermopower $S(T)$ measurements of high-quality YbRh$_2$Si$_2$ single crystals down to 30~mK. $S(T)$ is found negative with comparably large values in the paramagnetic state. In zero field no Landau-Fermi-liquid (LFL) like behavior is observed within the magnetically ordered phase. However, a sign change from negative to positive appears at lowest temperatures on the magnetic side of the QCP. For higher fields $B > B_c$ a linear extrapolation of $S$ to zero clearly evidences the recovery of LFL regime. The crossover temperature $T_{\rm LFL}$ is sharply determined and coincides perfectly with the one derived from resistivity $\rho(T)$ and specific heat $c_p(T)$ investigations.
\end{abstract}

\section{Introduction}
The appearance of quantum phase transitions at zero temperature and its consequences toward finite temperature have attracted considerable attention in the last years. The heavy-fermion compound YbRh$_2$Si$_2$ is regarded as a model system to study the influence of quantum criticality on the physical properties at low temperatures. It exhibits an antiferromagnetic (AFM) phase transition at an extremely low temperature of $T_{\rm N}$ = 70~mK. Upon applying a tiny magnetic field of $B_c$ = 60~mT within the tetragonal basal plane the AFM ordering is suppressed and the system is driven toward a field-induced quantum critical point (QCP) \cite{Gegenwart02}. Recently, an additional energy scale $T^*$ was found to vanish at the critical point in YbRh$_2$Si$_2$ \cite{Gegenwart07}, at which Hall effect measurements indicate a Fermi surface volume reconstruction according to the local moment scenario \cite{Si01}. 

The thermoelectric power is a sensitive tool to study low-energy excitations and can shed light on the evolution of the Fermi surface topology. For a degenerate Fermi gas within the Boltzmann picture and the relaxation time approximation the diffusion thermopower is given by the Mott formula $S_d = \pi^2k_B^2/(3e)T(\partial \ln \sigma/\partial \epsilon)_{\epsilon=\eta}$. The logarithmic derivative $(\partial \ln \sigma/\partial \epsilon)_{\epsilon=\eta}$ can be transformed into a transport term that contains the energy dependence of the scattering time and a term of purely thermodynamic character including the inverse of the effective mass tensor \cite{Miyake05,Behnia04}. Thus, thermopower reflects both the influence of the energy dependence of the density of states (DOS) and the quasiparticle scattering rate at the Fermi level. In the Landau-Fermi-liquid (LFL) state, when $(\partial \ln \sigma/\partial \epsilon)_{\epsilon=\eta}$ is temperature-independent, $S$ is proportional to $T$. In heavy-fermion metals due to the reduced characteristic energy scale $T_{\rm K}$ the renormalized electron masses and the ratio S/T are strongly enhanced. As a consequence, a close relation between the thermopower $S$ and the specific heat $c_p$ in the zero temperature limit is discussed \cite{Miyake05,Behnia04,Sakurai96,Zlatic07}.

Here, we report on low-temperature thermopower $S(T)$ measurements of a high-quality YbRh$_2$Si$_2$ single crystal down to 30~mK, tuning the ground state of the system from an antiferromagnetically ordered state to a non-magnetic Fermi-liquid state by the application of varying magnetic fields.

\section{Experimental Technique}
YbRh$_2$Si$_2$ single-crystalline platelets of high purity were grown from In flux with a molten-metal-solvent technique in closed Ta crucibles as described earlier \cite{Trovarelli00}. For the presented thermoelectric transport investigations a large, bar-shaped sample (4 $\cdot$ 0.5 $\cdot$ 0.1 mmm$^3$) with an optimized geometry factor of $l/A$ = 60~mm$^{-1}$ was chosen, whereas $l$ and $A$ represent the effective length and the cross section of the sample. The sample exhibited a residual resistivity $\rho_0$ = 2~$\mu\Omega$cm corresponding to a residual resistivity ratio $RRR \approx$ 40. 

The thermopower measurements in the temperature range 0.03~K $\leq T \leq$ 6~K were carried out in a commercial dilution refrigerator. The measurements were performed in steady-state conditions by utilizing a one-heater-two-thermometer-technique. The thermal gradient along the specimen $\Delta T = T_1 - T_2$ was determined by two RuO$_x$ chip resistors acting as hot and cold thermometer, while the heater in use was a meander-shaped resistor. The thermovoltage $\Delta U$ was measured with an analogue picovoltmeter by using superconducting NbTi wires connected to the sample. During experiment temperature gradients $\Delta T/\bar{T}\approx$ 0.01 were applied, with $\bar{T} = 0.5 \cdot (T_1 - T_2)$ being the mean temperature. The temperature stability of the two sample thermometers was usually better than 0.1\%. To achieve a good thermal contact the sample was tightly clamped to the cold finger and thermometry was connected to the sample by means of 50~$\mu$m Au wires. All thermometry equipment was suspended by thin nylon wires in a frame of vespel, electrical connections made with NbTi wires. The heat flow was always applied perpendicular to the $c$ axis of the tetragonal structure of YbRh$_2$Si$_2$. No respect was paid to the distinction between (100) and (110) direction within the $ab$ plane. Furthermore, the magnetic field was always parallel to the heat flow and therefore transverse effects in field are disregarded.

The reliability of the low-$T$ set-up was verified by measuring a Pb test sample below its superconducting transition temperature $T_c$ = 7.19~K. As the thermopower is zero in the superconducting state the results of this measurement are taken to determine the thermovoltage background signal $\Delta S$. It was found to scatter around zero with a magnitude of the order of 25~nV/K in the whole temperature range, only at lowest temperatures ($T<$ 200~mK) it increased to 50 nV/K.

\section{Results and Discussion}
For the analysis of the presented low-temperature data only the electron diffusion contribution $S_d$ to the thermopower has to be considered as it is described by the Mott formula (see above). A phonon-drag contribution $S^{\rm drag}_{\rm ph}$ to the thermopower usually yields a maximum in $S(T)$ at 0.1 - 0.2 $\Theta_D$ \cite{Paschen00}, with $\Theta_D$ being the Debye temperature of the system. As $\Theta_D$ was estimated to be of the order of 380~K \cite{Ferstl}, a possible phonon-drag contribution would not appear within the investigated temperature range and is therefore completely disregarded. Furthermore, crystal electric field (CEF) effects do not play any role in the low-$T$ range, as the excited levels of the Yb$^{3+}$ ion are well separated from the ground state \cite{Stockert06}. High-temperature thermopower studies on pure and Lu-doped YbRh$_2$Si$_2$ reveal the influence of the Kondo temperature $T_K$ at $\approx$ 10 - 20~K and the overall CEF splitting to appear at $\approx$ 80~K \cite{Hartmann06,Kohler08}, respectively. 

\begin{figure}
\begin{center}
\includegraphics[bb= 17 15 280 214,clip,width=0.65\textwidth]{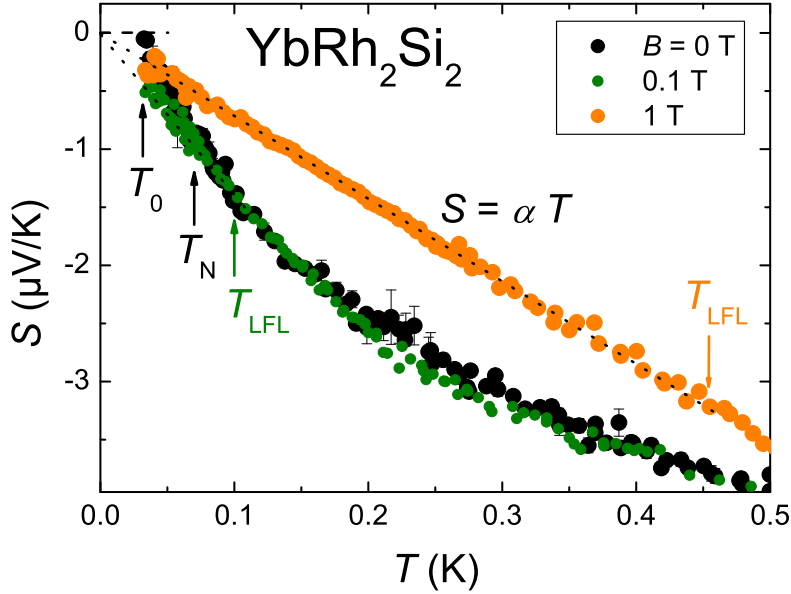}
\end{center}
\caption{\label{YbRh2Si2-S}Low-temperature thermopower $S$ of an YbRh2Si2 single crystal in various magnetic fields. For the zero field curve the Neel temperature $T_N$ = 70~mK and the temperature of the sign change $T_0$ are marked with arrows, as well as the temperature of Landau-Fermi-liquid recovery $T_{LFL}$ in magnetic field. The dotted lines represent fits to the data according to $S=\alpha T$ for $B$ = 0.1 and 1~T as described in the text.}
\end{figure}

Figure \ref{YbRh2Si2-S} shows the low-temperature part of the thermoelectric power $S(T)$ of an YbRh$_2$Si$_2$ single crystal for various magnetic fields applied. $S(T)$ exhibits a smoothly decreasing curvature with comparably large negative values over the whole investigated temperature and field range. High values and a negative sign are typically observed in Yb Kondo-lattice systems with the sharp Kondo resonance lying below the Fermi level. 

In zero field and at sufficiently low temperature the system is situated on the AFM side of the QCP, but very close to the magnetic instability. From thermopower measurements the phase transition at $T_{\rm N}$ = 70~mK is neither evidenced by any kind of anomaly, nor a distinct change of temperature behavior is resolved when entering the AFM phase. This is in contrast to observations from $c_{\rm p}(T)$ and $\rho(T)$ measurements, where $T_{\rm N}$ shows up as a sharp anomaly and a pronounced drop in zero field \cite{Gegenwart02}, respectively. However, a sign change of $S(T)$ from negative to positive appears in thermopower at $T_0 \approx$ 30~mK for $B$ = 0. This situation is very exceptional, as to our knowledge {\it positive} thermopower values at such low temperatures have never been observed in any Yb system before. The zero-field measurements do not resolve a linear-in-T behavior, neither above nor below $T_N$. Thus, in contrast to $c_{\rm p}(T)$ and $\rho(T)$ the thermopower results do not indicate a recovery of a LFL ground state within the magnetic phase.

Upon applying magnetic fields the ground-state properties of the system changes as it is driven toward the non-magnetic side of the QCP and LFL regime is recovered. This is evidenced by a $T^2$ behavior in resistivity and a constant Sommerfeld coefficient $\gamma$ \cite{Gegenwart02}. In the thermopower the low-temperature behavior strikingly changes compared to the zero-field curve. Below a certain temperature $T_{\rm LFL}$ the thermopower extrapolates linearly to zero for $T \rightarrow 0$, clearly evidencing the LFL regime from thermopower measurement. The temperature $T_{\rm LFL}$ below which LFL behavior is recovered is well distinguished and fully coincides with the findings from resistivity investigations on the same sample. With increasing field the range of LFL regime is extended to higher $T$, while absolute values are reduced and the slope of the curves changes. For temperatures well above 1~K all $S(T)$ curves finally merge and exhibit an almost identical $T$ dependence (not shown). The low-$T$ thermopower behavior within LFL phase is fitted according to $S = \alpha T$, with $\alpha$ being the initial slope of the curves. This yields large values of $\alpha$ = 13.5~$\mu$V/K$^2$ and 7.2~$\mu$V/K$^2$ for $B$ = 0.1 and 1~T, respectively. Enhanced values of $S/T$ have been reported for a few heavy-fermion systems (see for example \cite{Kuwai06}) and are considered to be in close relation to the strongly enhanced specific heat in these systems. When approaching the critical field $B_c$ of YbRh$_2$Si$_2$ the $A$-coefficient of the $T^2$ term to $\rho(T)$, the Sommerfeld coefficient $\gamma$ and the initial slope of $S$ are strongly increasing.

\begin{figure}
\begin{center}
\includegraphics[bb= 18 16 322 236,clip,width=0.65\textwidth]{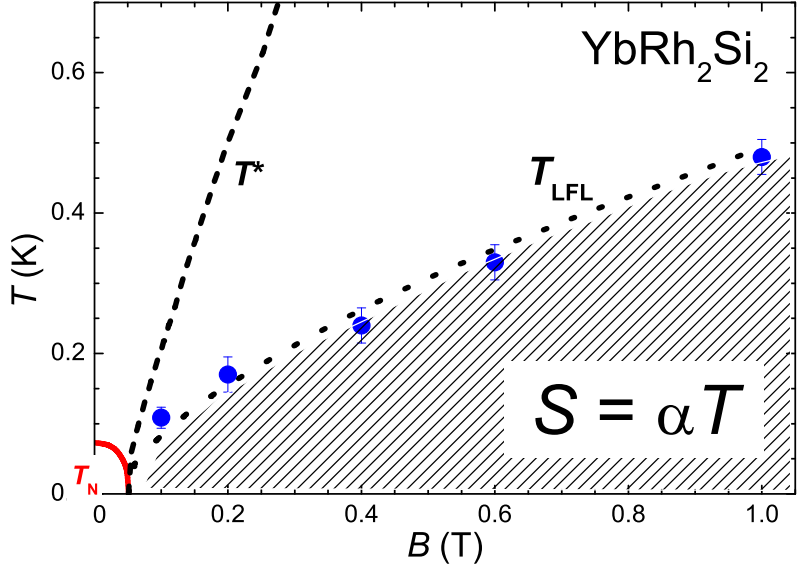}
\end{center}
\caption{\label{YbRh2Si2-phasediagram}Low-temperature $T-B$ - phase diagram of YbRh$_2$Si$_2$. The shaded region represents the Landau-Fermi-liquid phase, where the thermopower exhibits a linear temperature dependence as $S = \alpha T$, the blue symbols represent the temperature of LFL-recovery $T_{\rm LFL}$ from thermopower investigations compared to the observations in resistivity (dotted line). $T_N$ and $T^*$ lines are taken from \cite{Gegenwart07}.}
\end{figure}

Figure \ref{YbRh2Si2-phasediagram} shows the low-temperature $T-B$ - phase diagram of YbRh$_2$Si$_2$. The shaded region represents the non-magnetic LFL phase, where $S$ is proportional to $T$ together with a $T^2$ behavior in $\rho$ and $\gamma$ being a constant. The crossover line $T_{\rm LFL}$ as derived from resistivity measurements coincides perfectly with the recovery of LFL behavior as evidenced by the thermopower investigations.

%%%%%%%%%%%%%%%%%%%%%%%%%%%%%%%%%%%%%%%%%%%

\end{document}